# Imaginary Poynting momentum driven particle rotation by cylindrically polarized Gaussian beams

XUE YUN[1], YANSHENG LIANG[1,*], LINQUAN GUO[1], MINRU HE[1], TIANYU ZHAO[1], SHAOWEI WANG[1], MING LEI[1,2,*]

[1] MOE Key Laboratory for Nonequilibrium Synthesis and Modulation of Condensed Matter, Shaanxi Province Key Laboratory of Quantum Information and Quantum Optoelectronic Devices, School of Physics, Xi'an Jiaotong University, Xi'an 710049, China
[2] State Key Laboratory of Electrical Insulation and Power Equipment, Xi'an Jiaotong University. Xi'an 710049, China
yansheng.liang@mail.xjtu.edu.cn.
ming.lei@mail.xjtu.edu.cn

**Abstract:** Imaginary Poynting momentum (IPM) provides a new degree of freedom for particle manipulation. However, the application of IPM in experiments has been largely unexplored. Here, we demonstrate the IPM driven particle rotation by cylindrically polarized Gaussian beams with no spin or orbital angular momentum. Theoretical analysis and experimental measurements demonstrate that gold microparticles will be rotated in the azimuthal direction while confined in the radial direction. We achieved controllable rotation of the particle by tuning the cylindrical polarization state. Interestingly, the transfer of IPM to a gold particle is demonstrated to be competitive with that of spin angular momentum. These findings hold promising in light-matter interactions and particle manipulations.

1. Introduction

The technology for precisely rotating micro/nanoscopic particles has attracted tremendous scientific interest and holds substantial technical importance across various fields, including biomedicine, physics, and chemical engineering[1, 2]. For example, a rotated particle can serve as a probe to measure the viscous resistance of fluid [3-5]. The rotational dynamic of cells has been proven to be a key property for diagnosing the cell's viability[6, 7]. Additionally, the rotating bio-micromotors can be used for non-invasive targeted drug delivery within biological media [8, 9]. Among the numerous micromanipulation techniques[10-12], light-induced rotation is one of the most preferred methods owing to the merits of non-mechanical contact and low damage[13, 14]. The beams' angular momentum usually induces the torque that makes a particle rotate. In 1909, Poynting discovered that the circularly polarized beams carry spin angular momentum (SAM) of $\pm\hbar$ for each photon[15], which can be utilized to spin micro/nano-particles around its axis. Many years later, Allen et al. proposed that Laguerre-Gaussian beams with spiral phase term of $\exp(il\varphi)$ possess orbital angular momentum (OAM) of $l\hbar$ for each photon, with $l$ and $\varphi$ denoting the topological charge and the azimuthal coordinate. OAM generally drives particles to rotate along a designed trajectory[16]. Note that under specific conditions, strong coupling effects will occur between the SAM and OAM [17, 18]. For instance, a circularly polarized Gaussian beam will cause the orbital rotation of particles [19]. Given the common phenomenon caused by SAM and OAM, rotating particles without the transfer of angular momentum seems quite counterintuitive. However, Nieto-Vesperinas et al. demonstrated theoretically that the Poynting vector's imaginary part, usually considered a reactive quantity, can be transferred to matter to produce force[20]. Later, this highly extraordinary component was predicted to be observable in some physical pictures like evanescent fields or two-wave interference[21, 22]. Xu et al. indicated the presence of vortex-structured IPM density within paraxial cylindrically polarized beams, which could provide the force to drive objects to undergo orbital motion [23]. Despite the increasing theoretical frameworks, the experimental investigations of IPM are still in their infancy, especially in the research of optical tweezers. M. Antognozzi et al. observed the transverse spin-dependent motion in an evanescent field using a nano-cantilever [24]. However, such transverse motion is driven by the collective effect of the transverse Belinfante spin momentum force and the transverse IPM force. Recently, the mechanical effects of IPM were verified in the experiment using a cylindrically polarized perfect vortex beam[25]. Given the short in experimental exploration of IPM, the applications of utilizing IPM have never been reported. Therefore, the mechanisms and applications of IPM require further investigations, particularly in experimental research.

In this paper, we investigate the IPM driven particle rotation by tightly focused cylindrically polarized Gaussian beams. Theoretical simulations and experimental results indicate that in the absence of spin/orbital angular momentum, the azimuthal force arising from the transfer of IPM drives gold particles to rotate continuously along a circular orbit with the rotation rate dependent on the cylindrical polarization state. As the trapping position is far from the optical axis, the gravitational force acting on the gold particle is sufficient to counteract the strong scattering force of light, allowing the particles to rotate in three-dimensional (3D) space without a cover glass to balance the scattering force. Moreover, we demonstrate that IPM has a competitive transfer rate with SAM for tightly focused cylindrical polarized Gaussian beams. Our findings will facilitate a deeper understanding of the fundamental properties of IPM and pave the way for its application across various fields.

## 2. Principle of IPM-driven particle rotation

In this paper, we will investigate the mechanism of IPM-driven particle rotation in cylindrical polarized Gaussian beams under nonparaxial conditions. The Poynting vector, which describes the energy flux density in an electromagnetic field, can be expressed as:

$$\vec{\Pi} = \left(\vec{E}^* \times \vec{H}\right)/2 \tag{1}$$

Here, we consider a cylindrically polarized Gaussian beam illuminating a high numerical aperture (NA) objective lens with a focal length of $f$. The polarization direction at each point maintains a constant angle $\alpha$ concerning the radial direction (Fig. 1(a)-(c)). In cylindrical coordinates $(r, \phi, z)$, the electric field's components $(E_r, E_\phi, E_z)$ near the focal plane can be calculated by Richards-Wolf vectorial diffraction integral theory, written as

$$E_r(r,\phi,z) = -kf \cos\alpha \int_0^{\theta_{max}} E_0 \sqrt{\cos\theta} \sin\theta \cos\theta J_1(kr\sin\theta) \exp(ikz\cos\theta) d\theta, \tag{2}$$

$$E_\phi(r,\phi,z) = -kf \sin\alpha \int_0^{\theta_{max}} E_0 \sqrt{\cos\theta} \sin\theta J_1(kr\sin\theta) \exp(ikz\cos\theta) d\theta, \tag{3}$$

$$E_z(r,\phi,z) = -ikf \cos\alpha \int_0^{\theta_{max}} E_0 \sqrt{\cos\theta} \sin^2\theta J_0(kr\sin\theta) \exp(ikz\cos\theta) d\theta, \tag{4}$$

where $k$ is the wavenumber and $\theta_{max}$ denotes the maximal angle determined by the NA of the objective lens. $E_0$ is the electrical field of the incident Gaussian beam expressed as

$$E_0(\rho,\zeta,\eta) = \frac{\omega_0}{\omega(\eta)} \exp\left(-\frac{\rho^2 + \zeta^2}{\omega^2(\eta)}\right), \tag{5}$$

where $\omega_0$ is the waist width. To simplify the equation, we substitute the expression

$$U(r,z) = -kf \int_0^{\theta_{max}} E_0 \sqrt{\cos\theta} \sin\theta J_1(kr\sin\theta) \exp(ikz\cos\theta) d\theta, \tag{6}$$

into Eq. 2 - Eq. 4. Then we obtain the total electric field

$$\vec{E}(r,\phi,z) = -i\frac{\cos\alpha}{k}\partial_z U(r,z)\vec{e}_r + U(r,z)\vec{e}_\phi + i\frac{\cos\alpha}{k}(\partial_r + \frac{1}{r})U(r,z)\vec{e}_z. \tag{7}$$

From the Maxwell's equations, we have the following magnetic field:

$$\vec{H}(r,\phi,z) = i\frac{\sin\alpha}{\mu ck}\partial_z U(r,z)\vec{e}_r + \frac{1}{\mu c}\cos\alpha U(r,z)\vec{e}_\phi - i\frac{\cos\alpha}{\mu ck}(\partial_r + \frac{1}{r})U(r,z)\vec{e}_z. \tag{8}$$

Therefore, according to Eq. (1), Eq. (7), and Eq. (8), the real part of the complex Poynting vector is

$$\begin{aligned}\operatorname{Re}(\vec{\Pi}) &= \operatorname{Re}(\Pi_r) + \operatorname{Re}(\Pi_\phi) + \operatorname{Re}(\Pi_z) \\ &= \frac{1}{2\mu ck}\operatorname{Im}(U^* \cdot \partial_r U)\vec{e}_r + \frac{1}{2\mu ck}\operatorname{Im}(U^* \cdot \partial_z U)\vec{e}_z,\end{aligned} \tag{9}$$

This part represents time-averaged momentum density, with its angular component equal to zero. The IPM density is

$$\begin{aligned}\operatorname{Im}(\vec{\Pi}) &= \operatorname{Im}(\Pi_r) + \operatorname{Im}(\Pi_\phi) + \operatorname{Im}(\Pi_z) \\ &= \frac{\cos 2\alpha}{2\mu ck}\left(\frac{1}{r}|U|^2 + \frac{1}{2}\partial_r|U|^2\right)\vec{e}_r + \frac{\sin 2\alpha}{2\mu ck^2}\operatorname{Im}\left[\frac{1}{r}\partial_r(rU^*)\partial_z U\right]\vec{e}_\phi \\ &\quad + \frac{\cos 2\alpha}{2\mu ck}\operatorname{Re}(U^* \cdot \partial_z U)\vec{e}_z,\end{aligned} \tag{10}$$

where $\mu$ is the medium permeability and $c$ is the light speed in the medium. Equation (10) shows that the IPM density depends on the polarization angle $\alpha$. When $\alpha$ is equal to ±45°, the IPM has only the azimuthal component. Thus, particles in such a field will undergo rotational motion (Fig. 1(a) and (b)). At $\alpha = 0°$, the azimuthal IPM component decreases to zero. Thus, the particles will stop the rotation (Fig. 1 (c)).

## 3. Results

### 3.1 Orbital motion driven by IPM

To show the relationship between the IPM and the polarization angle $\alpha$ more clearly, in Fig. 2, we provide the simulated field distributions of the tightly focused cylindrically polarized Gaussian beams at the focal plane with $\alpha$=-45°, 45°, 22.5°, and 0°, respectively. Here. For simplicity, we set $E_0$=1, indicating that we use a plane wave with waist width $\omega_0$=∞. The NA for focusing was set to 1.33 for simulations. All the focal fields exhibit axially symmetric intensity patterns (Fig. 2(a)). The IPM flow (Fig. 2(b)) and the IPM components versus the radial position (Fig. 2c) indicate that, for $\alpha \neq 0$, the IPM distribution exhibits a vortex-like structure. In specific, for $|\alpha|$=45°, the azimuthal component Im($\Pi_\phi$) attains its maximum value, while the radial component Im($\Pi_r$) gets zero. When the polarization angle is in the opposite direction, the IPM vortex changes its orientation to the opposite direction. However, as $\alpha$ decreases to 0°, the azimuthal component decreases to zero, too, and the IPM has only the radial component with the maximal value. For the real part of the complex Poynting momentum, Re($\vec{\Pi}$), as shown in Fig. 2(d), there exists mainly the longitudinal part Re($\Pi_z$) despite the different polarization state. The radial component Re($\Pi_r$) is small and can be neglected (Fig. 2(e)). The azimuthal component Re($\Pi_\phi$) being zero indicates that the spin and orbital angular momentum are absent in the focused field.

To observe the rotational effect caused by the IPM of a cylindrically polarized Gaussian beam, we theoretically calculated the force experienced by a gold microparticle with a diameter of 4 μm in the x-z and x-y planes using the T-matrix method[26, 27]. The refractive index $n_{gold}$ of gold is 0.4 + 7.36$i$ at the wavelength of 1064 nm. We set the illumination power to 200 mW. The net force field considers the particle's gravity $F_g$=$\rho_{gold}gV_{gold}$, the buoyancy $F_f$=$\rho_{water}gV_{gold}$, and the optical force $F_o$, with $\rho_{gold}$ = 19.3 g/cm³ and $\rho_{water}$ = 1.0 g/cm³ respectively denoting the density of water and gold, $g$ = 9.8m/s⁻² being the gravitational constant, and $V_{gold}$ denoting the volume of the gold particle.

First, we validate the stable confinement of gold microparticles. The force map in the *x-z* plane for $\alpha$=−45° and a gold particle with a diameter of 4 μm is given in Fig. 3(a), showing two zero force points in regions 1 and 2, which are far away from the trap intensity center (Fig. 3(b)). To determine if the gold particle could be stably confined at these two positions, we computed the force vector maps (Figs. 3(c, d)) and the one-dimensional force profiles (Figs. 3(e, f)) in the vicinity of these two regions. The vortex-like force vector maps predict that the particle will be confined stably and transversely spun simultaneously, similar to the off-axis levitation of gold particles using linearly polarized Gaussian beams[28]. The transverse spin torque component is about −2.136 pN·μm. Further detailed analysis of the force profiles (Figs. 3(e, f)) confirms the stable trapping. Specifically, the equilibrium positions in regions 1 and 2 are at $(r, z)$=(±5.725, −0.10) μm. The far-away off-axis confinement allows rotation motion to be clearly observed.

Next, besides the transverse spinning motion, we will verify the orbital motion of the confined gold particle in tightly focused cylindrically polarized Gaussian beams. We conducted simulations for various cylindrical polarization angles of $\alpha$=−45°, 45°, 22.5°, and 0° (Fig. 4). All the net force distributions in the transverse plane (the first column in Fig. 4) present zero force positions (regions 1~4). The one-dimensional force profiles (the second and third columns in Fig. 4) confirm these positions as the equilibrium points along the radial direction, where the particle will undergo azimuthal driving forces $F_\phi$ for $\alpha \neq 0$, e.g., −45° (Fig. 4(a2-a4)), 45° (Fig. 4(b2-b4)), and 22.5° (Fig. 4(c2-c4)). The direction of the azimuthal force $F_\phi$ depends on the sign of the cylindrical polarization angle. Therefore, $F_\phi$ oriented along the positive direction for $\alpha$=−45° (Fig. 4(a4)), and negative for $\alpha$=45° (Fig. 4(b4)). When $\alpha$ decreases from 45° to 0, the azimuthal force at the radial equilibrium position decreases to 0 accordingly, while the radial force $F_r$ reaches the maximal value. These azimuthal and radial forces are induced by the transfer of the IPM. Therefore, the value and direction of the azimuthal and radial forces are related to the IPM distribution, as shown in the fifth column (the IPM density against the radial position) and sixth column (the corresponding IPM flow) of Fig. 4. It is noteworthy that for $\alpha$ = ±45°, the particle still experiences a radial force $F_r$ due to the intensity gradient of the light field, despite the radial component of the IPM being zero. This ensures that the particle is stably confined to the circular orbits due to radial restoring forces when the particle rotates.

## 3.2 Comparison of the IPM force and the spin-orbit coupling (SOC) force

Unlike the IPM driven rotation of particles, the SOC is a well-known resource that induces the orbital motion of particles [29]. Here, we attempt to investigate the comparison between these two types of light-induced particle orbital motion based on different mechanisms. We conducted simulations on the equilibrium positions and the azimuthal force for gold microparticles of various sizes and the azimuthal forces acting on the particles at these positions in cylindrically polarized and left-handed circularly polarized Gaussian beams ($\omega_0$=∞) of different powers. The equilibrium position profiles for cylindrical and circular polarization show similar trends with increasing illumination power and particle size. Specifically, the radial and axial equilibrium positions increase with illumination power (Fig. 5(a)). For increasing particle size, we obtain an increasing radial equilibrium position but a decreasing axial equilibrium position due to the increasing particle gravity (Fig. 5(b)). The azimuthal force is nearly unchanged with illumination power due to the changing equilibrium position (Fig. 5(c)), while it increases quasi-linearly with the particle size (Fig. 5(d)). Interestingly, we find that under the same focusing condition, the equilibrium position and the azimuthal force profiles for cylindrical polarization and circular polarization show high similarity, i.e., the equilibrium position and the azimuthal force are nearly the same at various illumination powers and for different particle size.

*3.3 Experimental result*

In the experiment, gold particles with a diameter ranging from 3.5 to 5.0 μm immersed in water were used as probes to verify the orbital motion induced by the IPM. Figure 6(a) shows the experiment setup. A linearly polarized laser with a wavelength of 1064 nm passes through a vortex half-wave retarder (WPV10L-1064, Thorlabs), converting the polarization state to a cylindrical polarization. Then, the beam is directed to a high-NA oil immersion objective (100 ×, NA 1.45, Nikon Inc) to form the optical trap. We used a camera (DCC3240M, Thorlabs) with a resolution of 1280 × 1024 pixels and a pixel pitch of 5.3 μm to monitor the manipulation process. Figure 6(b) provides the snapshots of a rotated gold microparticle ($d \approx 4.2$ μm) illuminated by a tightly focused cylindrically polarization Gaussian beam with $\alpha=45°$ and a power of 200 mW. Clearly, the particle is stably confined to a circular orbit with a radius of approximately 5.4 μm centered around the focal spot, rotating in a clockwise direction. The rotation rate and direction are related to the polarization angle $\alpha$. Detailed experimental measurements on the rotation rate and the simulated azimuthal force $F_\phi$ for various polarization angles are shown in Fig 6(c). As $\alpha$ varies from ±45° to 0°, the magnitude of the azimuthal force gradually decreases. When $\alpha$ is approximately ±15°, the azimuthal force is insufficient to overcome the viscous drag, causing the particles to stop rotating. Neglecting the cases of small polarization angle when the azimuthal force is too weak to induce the orbital motion, the experimentally measured rotation rate profile exhibits a decreasing trend with polarization angle, consistent with the theoretical prediction of the azimuthal optical force. Moreover, when the direction of the polarization angle is reversed, the direction of the azimuthal force also reverses, resulting in the particles rotating anticlockwise for $\alpha \in [-45°, 0°]$, and clockwise for $\alpha \in (0°, 45°]$. To further investigate the particle motion at different polarization angles, Fig. 6(d) shows the trajectories of the captured particles recorded at $\alpha$ of −45°, 45°, 25° and 0°, respectively. The first and second columns provide the particle's positions versus time in the x- and y-directions, respectively, and the third column shows a scatterplot of the particle's positions in the x-y plane. We find that for $\alpha = -45°, 45°$, and 25°, the particle undergoes orbital motion with different rates, while for $\alpha = 0°$, the particle remains stably confined in orbit due to the radial force.

## 4. Conclusion

In conclusion, we have demonstrated theoretically and experimentally the orbital motion of particles induced by the IPM of a tightly focused cylindrical polarized Gaussian beam. The extraordinary optical force arising from IPM enhances the transverse force acting on optically trapped particles. It differs from the usual optical pressure and gradient forces, providing a new degree of freedom for optically manipulating particles. From the theoretical and experimental results, we find that the IPM force is comparable to the SOC force under the same tight focusing condition, which serves as a new means to induce continuous orbital rotation of particles within a single optical trap. The direction of the orbital motion depends on the orientation of the polarization angle. This novel rotational capability could be used to realize biological machines within living cells or optically driven micropumps in microsystems.

**Funding**

This work was supported in part by the National Key Research and Development Program of China (2023YFF0722600, 2022YFF0712500); Natural Science Foundation of China (NSFC) (62135003, 62205267, 62205265); Natural Science Basic Research Program of Shaanxi (2022JZ-34, 2022JM-321, 2024JC-YBMS494); Shaanxi Fundamental Science Research Project for Mathematics and Physics (23JSY010); The Fundamental Research Funds for the Central Universities(xzy012023033).

**Disclosures**

The authors declare no conflicts of interest.

**Data availability**

Data underlying the results presented in this paper are not publicly available at this time but may be obtained from the authors upon reasonable request.

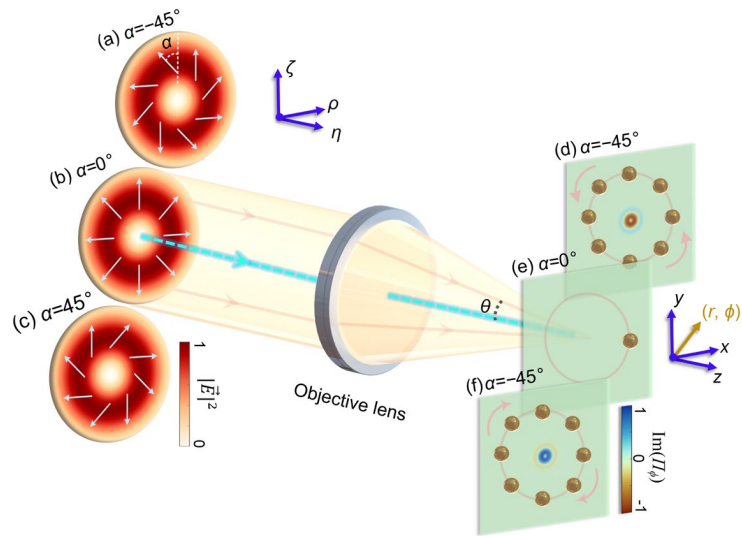

Fig. 1. Schematic illustration of the IPM driven particle rotation by a tightly focused cylindrically polarized Gaussian beam with different polarization angles. (a)–(c) Cylindrically polarized Gaussian beams beam with polarization angles $\alpha$=−45°, 0°, and 45°. (d)–(f) Corresponding IPM density distributions in the focal plane and schematic diagrams of the particle's orbital motions.

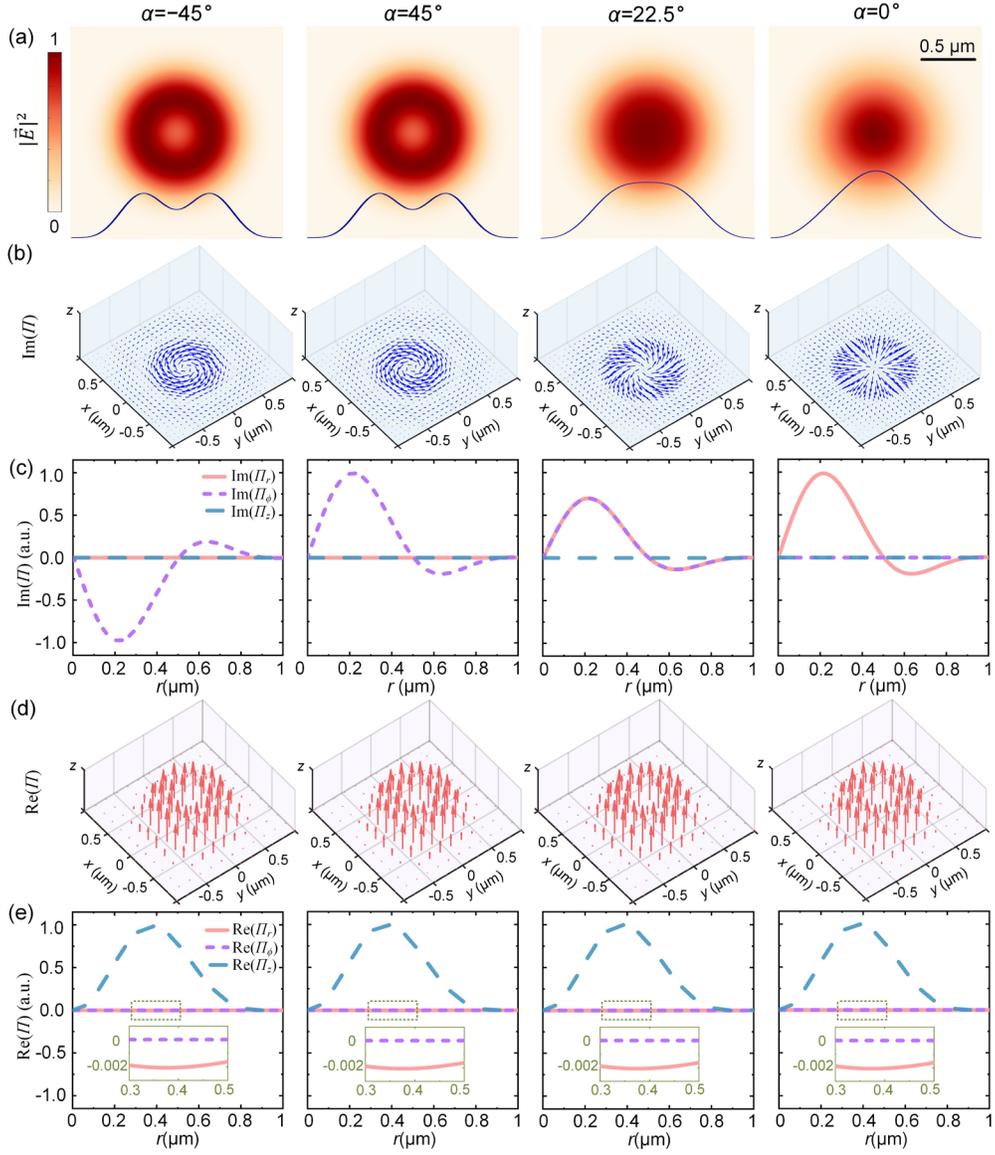

Fig. 2. The IPM distribution of tightly focused cylindrical polarized Gaussian beams with different polarization angles. (a) The intensity distribution. (b) IPM flow. (c) The radial, azimuthal, and axial components of the IPM related to the radial position. (d) The real part of the complex Poynting momentum Re($\Pi$). (e) The radial, azimuthal, and axial components of the Re($\Pi$) related to the radial position.

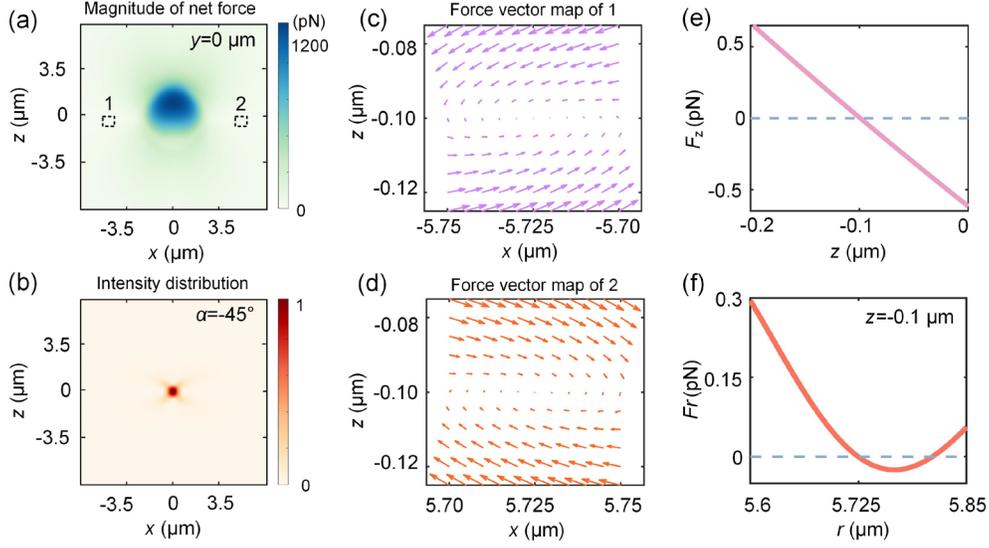

Fig. 3. Simulated optical force acting on a gold particle with a diameter of 4 μm in the *x-z* plane. The illuminating laser is a cylindrical polarized Gaussian beam with a polarization angle of α=−45° and an illumination power of 200 mW. (a) The magnitude of net force distribution at *y*=0 μm. (b) Corresponding light field intensity distribution in the axial plane. (c, d) 2D net force vector maps in regions 1 and 2 marked by black rectangular in (a). (e, f) Variation curves of axial force $F_z$ along the *z*-direction (e) and radial force $F_r$ along the radial position (f) in region 2.

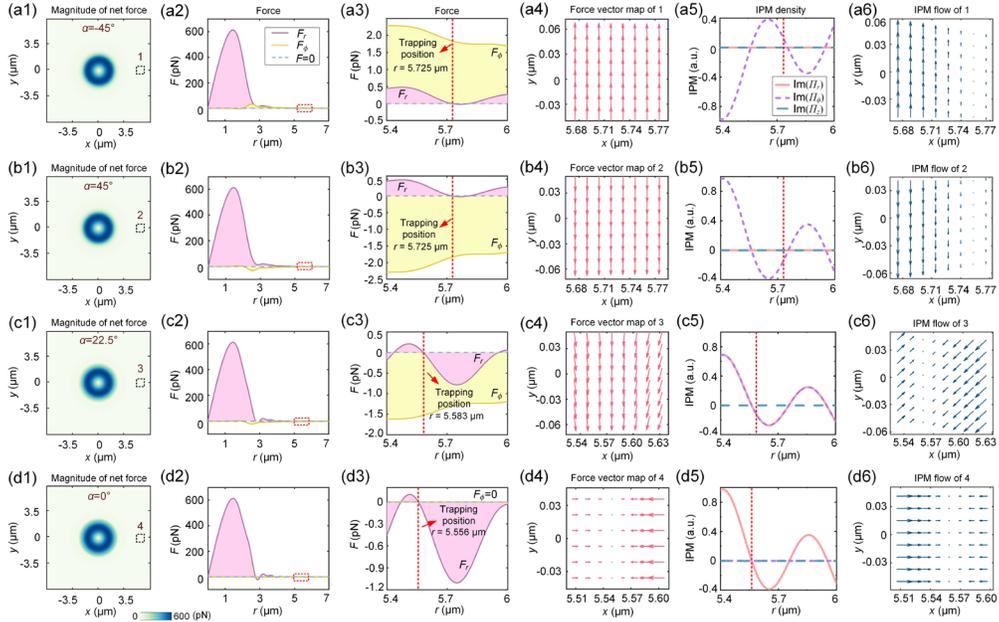

Fig. 4. Simulated optical force and IPM distribution in the transverse plane at different polarization angles α. (a) α=−45°. (b) α=45°. (c) α=22.5°. (d) α=0°. (a1) - (d1) The magnitude of net force distribution at *z*=−0.10 μm. (a2) - (d2) Radial force $F_r$ and azimuthal force $F_\phi$ versus the radial position. (a3) - (d3) Details in the red dotted boxs of (a2) - (d2). (a4) - (d4) 2D net force vector maps in regions marked by a black dotted box of (a1) - (c1). (a5) - (d5) The components of the IPM versus the radial position. (a6) - (d6) Corresponding IPM density.

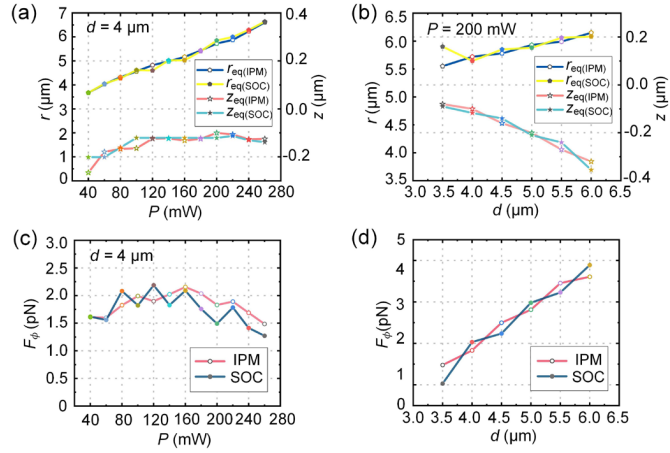

Fig. 5. Simulation of the particle trapping position as well as the magnitude of the azimuthal force $F_\phi$ when a strongly focused cylindrical polarized Gaussian beam with $\alpha=-45°$ and a left-handed circularly polarized (LCP) beam irradiate a gold particle. The calculated radial equilibrium position $r_{eq}$, axial equilibrium position $z_{eq}$ versus the power (a), and the diameter (b). (c) The azimuthal force acting on a 4 μm gold particle as a function of the illumination power. (d) The azimuthal force as a function of particle diameters when the power is 200 mW.

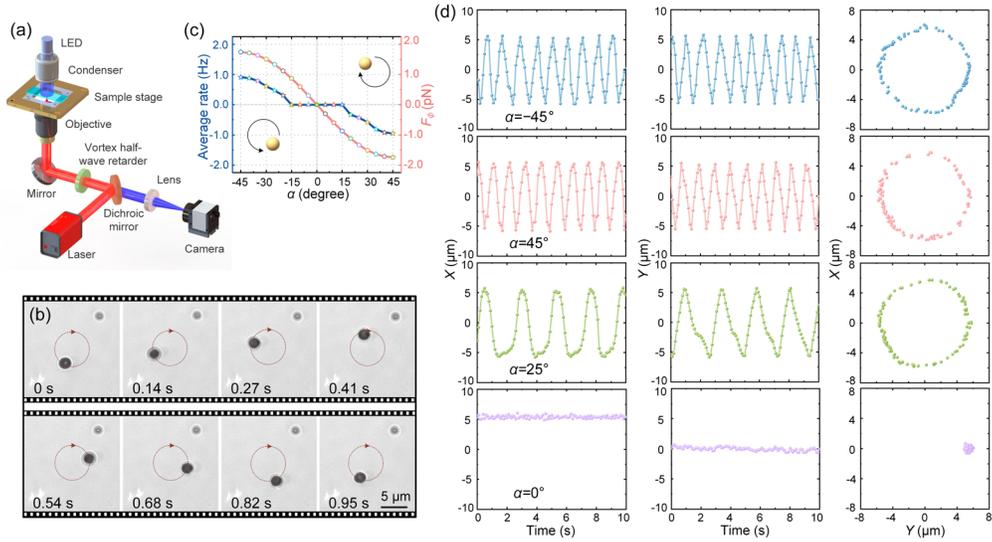

Fig. 6. Experimental results of rotating gold particle with a diameter of 4.2 μm by the tightly focused cylindrical polarized Gaussian beam. (a) Diagram of the experimental setup. (b) Snapshots of the rotated particle. (c) Simulations of the azimuthal force $F_\phi$ and the experimentally measured particle rotation rate for various polarization angles $\alpha$. (d) The trajectories of the trapped particle during 10 seconds.